# MODELISATION NUMERIQUE DE L'INTERACTION SOL-STRUCTURE LORS DU PHENOMENE DE FONTIS

# NUMERICAL MODELLING OF THE SOIL-STRUCTURE INTERACTION DURING A SINKHOLE

**Matthieu CAUDRON***
   INERIS, Parc technologique ALATA - BP 2, 60550 Verneuil-en-Halatte
   Phone : 00 33 (0)3 44 61 81 56; e-mail : matthieu.caudron@ineris.fr
**Fabrice EMERIAULT**
   LGCIE, INSA-Lyon, F-69621, France
   Phone : 00 33 (0)4 72 43 79 26; e-mail : fabrice.emeriault@insa-lyon.fr
**Marwan AL HEIB**
   INERIS, Ecole des Mines – Parc de Saurupt, 54042 Nancy Cedex
   Phone : 00 33 (0)3 83 58 42 97; e-mail : marwan.alheib@ineris.fr

**RESUME -** Cet article présente une étude du phénomène d'interaction sol-structure durant la formation d'un fontis par une modélisation numérique. L'approche utilise un modèle numérique bidimensionnel associant un code de calcul aux Différences Finies avec un code de calcul utilisant les Eléments Discrets afin de tirer au mieux parti des performances des deux logiciels. Outre un important gain de temps de calcul par rapport à un modèle constitué uniquement d'éléments distincts, cette approche donne des résultats comparables à ceux observés expérimentalement auparavant.

**ABSTRACT –** This article focuses on the simulation of soil-structure interaction during a sinkhole development by the use of a coupling numerical modelling approach. The 2D model uses a Finite Difference computer code associated with a Distinct Elements code to optimize the performances of both softwares. This allows an important decrease of computation time and the results computed are close of the experimental observations made before.

Mots-clés : fontis, interaction sol-structure, modèle numérique, méthode des éléments distincts

# 1. Introduction

Les fontis sont causés par l'effondrement de cavités souterraines peu profondes d'origine anthropique ou naturelle. On en recense un nombre important sur l'ensemble du territoire français. Ces cavités, du fait du caractère inéluctable de leur dégradation pourront s'effondrer (dégradation des propriétés des matériaux dans le temps, influence des variations des conditions environnementales telles que la température, l'hygrométrie, …) et donneront ainsi naissance à des fontis (figure 1). La nature brutale de ce phénomène peut être très préjudiciable pour les structures et infrastructures en surface ainsi que pour la population (Pacyna et al. 2005). A cause de leur large répartition géographique, de nombreuses communes et collectivités territoriales sont concernées par les conséquences de ces aléas sur les ouvrages (bâti, infrastructures, ...).

L'INERIS travaille depuis plusieurs années (Abbass-Fayad 2004) à "l'Analyse, la Prévention et la Maîtrise des risques de mouvements de terrains liés à la présence de cavité souterraines", avec comme objectif l'évaluation et le perfectionnement des outils permettant d'améliorer la gestion des risques liés aux cavités souterraines. Contrairement aux cas des travaux souterrains (tunnels, cavités profondes en site urbain, ...) (Burland & Wroth 1974; Burland 1995 ; Franzius et al. 2004 ), le risque encouru par les structures lors de mouvements de terrain résultant d'effondrements de cavités est moins bien connu.

Cet article s'attache plus particulièrement à la conception et à l'utilisation d'un modèle numérique alliant une approche de type Milieu Continu avec une approche de type Milieu Discontinu permettant d'étudier les phénomènes d'interaction sol-structure lors de l'apparition en surface d'un fontis (Caudron et al. 2006b). En effet, l'approche communément employée lorsqu'il s'agit d'étudier l'impact d'un fontis (ou de mouvements de terrain divers) sur une structure en surface, tend à dissocier les deux parties du problème. Les mouvements de terrain sont déterminés dans une première partie ne prenant pas en compte la présence du bâtiment : c'est une étude en condition de terrain vierge. Puis, dans un second temps, les déplacements et déformations déterminés précédemment sont appliqués à la structure en considérant par exemple les recommandations émises par l'AFTES (1995). Les phénomènes d'interaction sol-structure dus aux mouvements de terrain peuvent modifier les sollicitations transmises par le sol à la structure et réciproquement (Augarde et al. 1998 ; Vaillant et al. 2005).

Le modèle numérique présenté par la suite permet de prendre en compte de manière explicite l'interaction sol-structure. Ainsi, par comparaison entre des résultats obtenus en condition de terrain vierge puis en présence d'un modèle de bâtiment, il met clairement en évidence l'influence des interactions sol-structure lors du phénomène de fontis.

Le modèle fait appel à une association de deux logiciels (FLAC$^{2D}$ et PFC$^{2D}$) basés sur des méthodes numériques différentes ; une comparaison pertinente est possible entre simulations numériques et résultats d'une campagne expérimentale menée sur un modèle réduit de laboratoire (Figure 2).

Après avoir brièvement détaillé le cas d'étude et le modèle réduit de laboratoire qui a été utilisé et présenté dans Caudron et al. (2006a), le schéma de couplage utilisé entre les deux codes de calcul sera détaillé. Ensuite les résultats de simulation en condition de terrain vierge seront abordés, en terme de phénomène, de répétabilité et de déplacements en surface. Pour finir, des simulations en présence d'un modèle simplifié de structure en surface seront présentées.

## 2. Cas étudié

Le modèle réduit expérimental représente, à une échelle 1/40ème, un massif de sol d'une largeur de 30 m et de 12 m de hauteur de recouvrement au dessus d'une cavité (soit un modèle réduit de 750 mm par 300 mm). Ce recouvrement est constitué de deux couches différentes. Un banc résistant présentant une épaisseur équivalente à 2 m constitue le toit de la cavité (dans le modèle expérimental, ce matériau est représenté par des rouleaux de Schneebeli, dont un usage récent a été réalisé par Dolzhenko et al. (2001), rendus cohérents par ajout de colle aqueuse (Caudron et al. 2006a)). Ce banc cohérent est nécessaire pour permettre la création d'une cavité stable de taille supérieure à quelques rouleaux dont l'effondrement sera provoqué ultérieurement. Il est surmonté par une couche équivalente à 10 m de matériau granulaire frottant (représenté par le matériau de Schneebeli standard). Les caractéristiques des deux matériaux sont présentées dans le tableau I. Le module de Young varie dans la plage de 50 à 100 MPa en fonction du confinement utilisé.

Tableau I. Caractéristiques mécaniques des deux matériaux.

Table I. Mechanical characteristics of the two materials.

|  | Densité | E (MPa) | $\varphi$ (°) | c (kPa) |
|---|---|---|---|---|
| Sol pulvérulent (réel) | 2.2 | 50-100 | 22-24 | 0 |
| Sol pulvérulent (modèle) | 2.2 | 50-100 | 24 | 0 |
| Sol cohérent (réel) | 2.2 | 50-100 | 27-30 | ~100 |
| Sol cohérent (modèle) | 2.2 | 50-100 | 25-30 | 65-150 |

La cavité, qui correspond à une hauteur équivalente de 2 m, est créée progressivement jusqu'à une largeur maximale correspondant à 10 m en 5 étapes, grâce à un appareillage inspiré des études menées par Nakai et al. (1997) sur les mouvements de terrain relatif au creusement de tunnels. La cavité centrale est ouverte par déplacement de parties mobiles en perturbant le moins possible le massif de sol. Puis à chaque étape supplémentaire, 1 m de part et d'autre sont ouverts de la même manière, l'ensemble est résumé par la figure 3.

La structure placée en surface est de type "poutres-poteaux" en acier. Ses caractéristiques géométriques sont détaillées dans le tableau II. Le chargement appliqué suit les règles de l'art (réglementation française : Fascicule 62 titre V).

Tableau II. Caractéristiques géométriques de la structure (taille équivalente).

Table II. Geometric characteristics of the building (corresponding size).

| Largeur de semelle | 1.65 m |
|---|---|
| Longueur des poutres | 3.2 m |
| Hauteur d'un étage | 2.7 m |
| Vide sanitaire | 1.2 m |
| Section (m²/ml) | 0.051 m² |
| Inertie (m$^4$/ml) | 60.8 10$^{-4}$ m$^4$ |
| Nombre de travées | 3 |

## 3. Le modèle couplé

### 1) But du couplage

Le modèle numérique utilisé tire parti des possibilités de couplage présentes entre les logiciels FLAC$^{2D}$ et PFC$^{2D}$. Cela permet de différencier la modélisation en l'adaptant au mieux pour un problème donné. Nous pouvons ainsi utiliser les Eléments Distincts pour représenter le massif de sol constituant le toit de la cavité, zone où se produiront les déplacements les plus importants ainsi que l'apparition de fractures au sein du matériau tandis que le reste du massif de sol (où n'apparaîtront que de faibles déplacements et déformations) et le modèle de bâtiment seront modélisés par FLAC$^{2D}$ (basé sur une approche en Différences Finies) en utilisant respectivement des éléments volumiques et des éléments de structures. La figure 4 représente schématiquement le raisonnement utilisé.

L'intérêt de cette approche est de bénéficier au maximum des avantages de chacun des deux codes de calcul, tout en limitant leurs inconvénients. Ainsi, la capacité intrinsèque de PFC$^{2D}$ à représenter la chute de blocs de matériau, ainsi que l'apparition de fractures sans intervention de la part de l'utilisateur, ni de définition préalable de directions ou de localisations particulières est intéressante pour le recouvrement situé à l'aplomb de la cavité. Par ailleurs, l'utilisation de FLAC$^{2D}$ pour les zones concernées uniquement par de faibles déformations permet un gain de temps de calcul non négligeable par rapport à un modèle équivalent basé sur PFC$^{2D}$ uniquement.

### 2) Présentation de FLAC$^{2D}$ et PFC$^{2D}$

FLAC$^{2D}$ est un logiciel de modélisation numérique utilisant la méthode des Différences Finies appliquée à un milieu continu. Il est extrêmement souple en terme.

PFC$^{2D}$ est également commercialisé par la société Itasca. L'originalité de son approche provient du fait qu'il utilise la Méthode des Eléments Distincts (MED) (Itasca Consulting Group, 2005). Il considère ainsi le sol comme un ensemble de particules circulaires indéformables possédant chacune ses caractéristiques propres, ce qui lui confère la possibilité de mieux représenter le comportement réel d'un sol soumis à de grandes déformations et à de la fracturation qu'une approche plus classique de type milieu continu. Cependant cette approche présente deux inconvénients ou limitations majeurs :

- Le temps processeur requis pour modéliser un même problème est beaucoup plus important que pour les logiciels utilisant une approche en milieu continu. Cela dépend cependant beaucoup de la loi de comportement et du schéma de résolution retenu.

- Le comportement global d'un ensemble de particules est la résultante de l'ensemble des propriétés des particules. Le comportement macroscopique est étroitement lié aux propriétés des contacts interparticulaires des particules. Une délicate phase de calage des paramètres est donc nécessaire car il n'existe pas de relations directes entre les micropropriétés et les caractéristiques macroscopiques usuelles des sols: $\varphi$, c, E, $\psi$…

Le calage des différents paramètres de contact interparticulaire des particules est réalisé au moyen d'essais biaxiaux. Les paramètres sont ajustés jusqu'à obtenir le comportement du sol souhaité présenté dans le tableau I.

Enfin, chacun des deux logiciels peut "communiquer" avec une autre application développée par Itasca à travers une connection réseau. Ceci permet donc aux deux logiciels d'échanger les données nécessaires au bon fonctionnement du couplage (forces et déplacements au niveau de la zone frontière) de manière simple.

### 3) *Fonctionnement du couplage*

Le maillage FLAC$^{2D}$ désiré est créé, de même pour l'assemblage de particules PFC$^{2D}$ et un ensemble de procédures va échanger entre les deux codes les informations nécessaires au niveau des frontières communes. Plus particulièrement, PFC$^{2D}$ transmet à FLAC$^{2D}$ les forces qu'appliquent les particules sur les arêtes de FLAC$^{2D}$ concernées et FLAC$^{2D}$ renverra les vecteurs vitesses d'un certain nombre de particules asservies car en contact avec les frontières de FLAC$^{2D}$.

Ce schéma (Figure 5) est valable pour un cas simple où il n'y a pas de perte ou d'apparition de contact entre des particules et des éléments frontières. Or dans un modèle représentant la formation d'un fontis, il est nécessaire de pouvoir déterminer les pertes de contact lors de la création de la cavité puis les créations de contact lors de la rupture du toit de la cavité et de la chute des particules à l'intérieur de celle-ci. Pour ce faire, nous avons mis au point une deuxième procédure qui englobe le précédent schéma de fonctionnement.

Il est nécessaire de déterminer tous les *x* cycles de calcul s'il y a apparition de nouveaux contacts entre des particules et des arêtes frontières. Ce nombre de cycles maximal est déterminé à partir de la vitesse maximale des particules à un instant donné. Il sera tel qu'une particule ne pourra pas se déplacer de manière trop importante durant ces *x* cycles avant de déterminer à nouveau les particules en contact avec des éléments frontières. La distance limite est prise égale au rayon de la plus petite particule ($R_{min}$ = 3mm dans notre cas). La procédure suivante est ainsi appliquée:

1. L'utilisateur spécifie un nombre total de cycles de calcul à réaliser *X*.
2. Une fonction Fish (langage de programmation propre aux logiciels Itasca) détermine la vitesse maximale $V_{max}$ sur l'ensemble des particules. Le programme détermine alors le nombre maximum de cycles qui peuvent être réalisés en toute sécurité avec
$$x = R_{min}/V_{max}$$
3. Le modèle couplé réalise *y* cycles de calcul en utilisant le schéma de couplage détaillé précédemment, *y* étant le plus faible des deux nombres *x* et *X*.
4. PFC$^{2D}$ libère les particules asservies, reçoit de nouveau la liste des éléments frontières de FLAC$^{2D}$ et détermine une liste de nouvelles particules asservies.
5. On recommence à l'étape 2 en remplaçant *X* par *X-y*, jusqu'à ce que le nombre de cycles de calcul effectif soit égal au nombre de cycles demandé par l'utilisateur.

Si l'on souhaite modifier le maillage FLAC$^{2D}$ ou la géométrie de la frontière entre FLAC$^{2D}$ et PFC$^{2D}$ (dans notre cas particulier pour créer la cavité à l'origine du fontis), on suit la même procédure en effectuant les modifications et la mise à jour entre les étapes 3 et 4.

Dans le cas d'une structure en surface, celle-ci est modélisée par FLAC$^{2D}$. Par ailleurs, l'interfaçage entre éléments de structure et particules PFC$^{2D}$ n'étant pas possible à l'heure actuelle directement par l'intermédiaire des procédures de couplage Itasca, il a été choisi "d'enrober" les semelles dans des zones FLAC$^{2D}$ qui peuvent interagir avec les particules PFC$^{2D}$. Des zones de petites dimensions "flottantes" sont donc créées et rendues solidaires des semelles de la structure localisées dans la région représentée par PFC$^{2D}$ (Figure 6). Ces zones présentent une densité de maillage très faible (2*2 mailles). Cela est dû à la nécessité de conserver une taille d'élément supérieure à plusieurs fois la taille moyenne d'une particule PFC.

### 4) Calage des paramètres

Le calage des paramètres de contact interparticulaire de PFC$^{2D}$ est basé sur les résultats de plusieurs essais biaxiaux (courbes $q - \varepsilon_1$ et $\varepsilon_v - \varepsilon_1$). Les différents paramètres régissant le comportement local des particules sont les suivants:

- kn et ks, raideurs normale et tangentielles du contact entre deux particules;
- fric, le coefficient de frottement intergranulaire;
- les paramètres c_n et c_s, limites de résistance normale et tangentielle du contact de type "Contact Bond" utilisé pour représenter la cohésion d'un sol dur ou d'une roche (Potyondy & Cundall, 2004).

Deux hypothèses simplificatrices (kn / ks = 2 et c_n / c_s = 1) permettent de ramener le problème au calage de trois paramètres indépendants. Une première série d'essais (Figures 7 et 8) permet de déterminer le meilleur jeu de paramètres pour le sol pulvérulent, une deuxième est nécessaire pour le matériau cohérent (tableau III). Des différences sont observables entre le comportement du matériau expérimental et le modèle numérique obtenu. Le modèle numérique est, en début d'essai, légèrement plus raide que les résultats expérimentaux. Le comportement volumique présente lui une différence importante à l'origine qui s'estompe vers la fin de l'essai. En effet, le matériau de Schneebeli est plus rapidement dilatant alors que l'assemblage de particules PFC$^{2D}$ passe par une phase initiale de contractance avant de présenter un comportement dilatant. Cependant les angles de dilatance obtenus sont sensiblement équivalents.

Tableau III. Caractéristiques mécaniques et paramètres des lois de contact dans PFC$^{2D}$.

Table III. Mechanical characteristics and contact law parameters for PFC$^{2D}$.

| Caractéristiques mécaniques | Densité | E (MPa) | φ (°) | c (kPa) |
|---|---|---|---|---|
| Sol pulvérulent (PFC) | 2.2 | 50-100 | 26 | 0 |
| Sol cohérent (PFC) | 2.2 | 100-150 | 30 | ~84 |
| **Paramètres** | $\rho$ (kg/m$^3$) | $k_n$ (N/m) | $\eta$ (°) | $c_n$ (N) |
| Sol pulvérulent | 2600 | 22.10$^6$ | 42 | 0 |
| Sol cohérent | 2600 | 22.10$^6$ | 16,7 | 100 |

## 4. Résultats des différentes simulations

Deux cas de simulations numériques d'effondrement de cavité ont pu être simulés. Le premier est en condition de terrain vierge, c'est à dire en l'absence de structure, et sert de référence. Le second cas prend en compte la présence de la structure afin de déterminer l'importance du phénomène d'interaction sol-structure sur la formation du fontis.

La première configuration a fait l'objet de cinq calculs utilisant un assemblage initial différent des particules constitutives du massif. La seconde configuration a été simulée deux fois (seul l'assemblage des particules variant entre ces deux simulations), de la même manière que pour les simulations en condition de terrain vierge.

Les résultats qui seront plus particulièrement analysés, sont l'amplitude maximale de déplacement en surface et sa localisation, la forme de la cuvette d'affaissement, la pente sous l'emprise de la structure. Ces différents indicateurs sont en effet caractéristiques des sollicitations appliquées à un bâtiment en surface lors de mouvements de sol et permettent de définir le niveau d'endommagement plausible (Deck et al., 2003).

### 1) Résultats en condition de terrain vierge

La cavité est créée en 5 étapes. La rupture apparaît, pour chacune des cinq simulations, lors de la cinquième étape, sans aucune intervention de l'utilisateur (c'est-à-dire sans fragilisation du banc raide au milieu du toit réalisée par l'expérimentateur). C'est donc une rupture purement mécanique, la résistance à la traction $C_n$ ou au cisaillement $C_s$ des contacts interparticulaires étant atteinte.

La figure 9 illustre le réseau de forces de contact intergranulaires dans le massif de sol lorsque la cavité est à la quatrième étape de sa création pour la simulation n°4. Le report de charge, de part et d'autre de la partie centrale de la cavité, qui se développe au sein du banc de matériau résistant est clairement visible. On observe de même la présence de contraintes de traction dans le banc raide, en zone inférieure de la partie centrale de la cavité, en zone supérieure à l'aplomb des parois latérales de la cavité.

Les reports de charge qui s'effectuent durant la création de la cavité sur les éléments-frontières de FLAC$^{2D}$, sont tracés sur la figure 10a pour la simulation n°4. Le comportement obtenu est très proche des valeurs "théoriques" attendues (en supposant que 90% du poids des terrains supportés par les cales que l'on abaisse se reporte sur les deux cales voisines), tracées sur la figure 10b.

La figure 11 montre l'état du modèle après rupture de la cavité. Il apparaît qu'un certain volume de vide subsiste, ceci étant dû au réarrangement des blocs lors de leur chute. Une cuvette d'affaissement en surface est clairement identifiable. Elle est reportée sur la figure 12.

La courbe "théorique" présentée sur cette même figure correspond à l'utilisation d'une version modifiée de l'approche empirique de Peck (1969) proposée par Caudron et al. (2006a) dans le cas de cavités de forme rectangulaire. Elle prend en compte une estimation du foisonnement pour une constitution des bancs de recouvrement connue.

Le volume de la cuvette est déterminé à partir du volume de la cavité affecté d'un facteur traduisant la propension au foisonnement.

$$S_x = S_{max} \times e^{\left(\frac{-x^2}{2i^2}\right)}$$

avec $S_{max}$ l'affaissement maximal, $i$ la distance au point d'inflexion et $x$ la position par rapport au centre de la cuvette.

De même, les déplacements horizontaux en surface peuvent être estimés au moyen d'une approche empirique proposée par Lake et al. (1992).

$$v_x = S_x \frac{(x/x_0)}{(Z_0/x_0)^\beta}$$

avec $x_0$ valant 1m, $Z_0$ la distance entre le centre de la cavité et la surface, β un paramètre permettant le calage de l'expression par rapport aux caractéristiques du recouvrement (il vaut dans notre cas 0,87).

La figure 13 présente les déplacements horizontaux obtenus en surface pour les cinq simulations ainsi que pour la campagne d'essais expérimentaux réalisée antérieurement.

Un certain nombre de différences sont observables entre les résultats. Au niveau de la cuvette d'affaissement, la comparaison est globalement satisfaisante, tant au niveau de l'affaissement maximal que de la pente maximale située au niveau du point d'inflexion. On note cependant que les valeurs obtenues au niveau de la zone centrale de la cuvette sont sensiblement plus importantes avec un écart de 25 à 30% de la valeur expérimentale, portant l'affaissement maximal à près de 34 mm, soit un équivalent de 1,36 m en vraie grandeur.

Les déplacements horizontaux sont par contre plus faibles pour les valeurs calculées par la modélisation numérique, avec une différence pouvant atteindre 65%. L'origine de cette différence, située dans la modélisation effectuée avec FLAC$^{2D}$ sera détaillée ultérieurement.

La variabilité des affaissements, due aux différences d'agencements des rouleaux, observés expérimentalement lors de la rupture de la cavité est reproduite numériquement. Par contre, celle des déplacements horizontaux est plus faible, voisine de ±2 mm. On observe de même une discontinuité importante entre les parties modélisées par FLAC$^{2D}$ et celle représentée par PFC$^{2D}$, que ce soit au niveau des déplacements verticaux qu'horizontaux. Nous y reviendrons dans la suite de cet article, lors de la présentation de l'origine de certaines différences dans les résultats de la modélisation numérique.

En terme de sollicitations qui seraient transmises à une structure en surface, on retiendra une pente maximale de l'ordre de 17%, soit une valeur sensiblement plus importante que la valeur de 12,3% observée expérimentalement. Ceci concorde néanmoins avec l'augmentation de la valeur de l'affaissement maximal.

La déformation horizontale est calculée comme étant le déplacement relatif entre des particules en surface. Pour limiter la variabilité inhérente à la formulation en éléments distincts, un pas glissant de calcul valant 40 mm (soit la largeur d'une semelle de la structure, correspondant à 1,60 m en pleine échelle) est utilisé entre deux particules considérées. La déformation horizontale varie alors entre +7% et -12,5% selon la localisation, ce qui est légèrement plus faible que les valeurs correspondantes obtenues expérimentalement : de +8% à -17%.

## 2) *Origine des différences de résultats entre les essais expérimentaux et les simulations numériques*

Comme on peut le constater sur les figures 12 et 13, les deux zones périphériques modélisées par FLAC$^{2D}$ se déforment très peu par rapport aux valeurs observées expérimentalement. Ceci occasionne une discontinuité importante visible sur les courbes d'affaissement vertical et de déplacement horizontal. Deux causes sont imputables :

- le maillage utilisé en surface est trop grossier par rapport aux déformations attendues. Aucune plastification ne peut apparaître, le comportement est donc limité à de l'élasticité linéaire.
- En l'absence de plasticité, le module de Young utilisé en surface est donc sur-évalué par rapport au comportement réel du matériau.

Une première solution serait de raffiner le maillage à ce niveau afin de limiter l'impact de celui-ci sur le résultat. Ceci est difficilement réalisable car il est important de conserver une taille d'élément de maillage FLAC$^{2D}$ supérieure à plusieurs fois la taille moyenne d'une particule de PFC$^{2D}$.

Une autre solution serait de diminuer le module de Young utilisé pour compenser le fait que de la plastification ne puisse pas se développer. D'autres tests seraient alors nécessaires afin de s'assurer de la pertinence de cette approche, qui pourrait perturber la transmission de l'énergie entre les deux parties du modèle.

### 3) Résultats en présence de la structure

La simulation numérique prenant en compte une structure en surface est illustrée sur la figure 14. La rupture obtenue lors d'une des simulations est différente de celle observée en condition de terrain vierge.

Lors de la première simulation, la rupture est obtenue durant la dernière étape de création de la cavité par ruine purement mécanique du banc raide. Par contre, lors de la seconde simulation, réalisée dans des conditions identiques à la première, excepté l'arrangement initial des particules, la cavité est stable après la dernière étape de sa création. Il a été nécessaire d'initier la dégradation pour obtenir la rupture du banc résistant.

Cette dégradation a été induite par l'intermédiaire d'une réduction de la valeur limite des efforts repris par les liaisonnements au sein du banc de matériau cohérent. La zone affectée par ce procédé est localisée au centre du toit de la cavité et remonte à travers le banc raide jusqu'à l'obtention de la rupture. Celle-ci se produit sous la forme de deux blocs de matériau cohérent qui tombent dans la cavité comme illustré par la figure 15.

Le sol pulvérulent en surface suit leur mouvement, les fondations de la structure se trouvent alors sollicitées. L'état stable final qui est atteint présente deux vides importants qui subsistent au niveau de la cavité et quelques vides de moindre ampleur au niveau des zones de rupture dans le banc raide.

La figure 16 montre la cuvette d'affaissement obtenue après stabilisation de la rupture. Le déplacement des semelles de la structure est aussi tracé. On remarque une différence entre les mouvements du sol et ceux des semelles. Elle est due à la solution technique utilisée pour effectuer le couplage entre les éléments de structure FLAC$^{2D}$ et les particules de PFC$^{2D}$. Le fait d'utiliser des zones volumiques FLAC$^{2D}$ perturbe de manière importante le comportement des semelles par rapport à ce qui est observé lors des essais expérimentaux. On retrouve de plus des différences entre les résultats des deux simulations, provenant de la différence de rupture obtenue au niveau du banc résistant. Le volume de matériau tombé diffère donc, de même que le volume de la cuvette d'affaissement en surface.

L'affaissement maximal observé est de 22 mm environ pour les semelles (0,88 m en vraie grandeur), ce qui représente une réduction importante par rapport au résultat obtenu en terrain vierge. Il en est de même pour la pente maximale simulée sous la structure qui est de 11%, soit une réduction de 6%. Ces variations sont à l'opposé de ce qui est observé lors d'essais similaires sur le modèle réduit physique, où une augmentation de l'affaissement maximal et de la pente est obtenue. La raison de cette différence est inhérente au comportement du modèle FLAC$^{2D}$, qui ne reproduit pas totalement le comportement observé expérimentalement.

Les déplacements horizontaux, tracés sur la figure 17, présentent moins de différences entre les résultats des deux simulations en présence de la structure. Par ailleurs, on obtient une diminution des déformations horizontales extrêmes au niveau de la structure par rapport aux simulations en terrain vierge (+1,9% à -0,9% par rapport à +6% et -13% en terrain vierge). Ceci rejoint les observations faites lors des essais expérimentaux.

### 4) Déformations au sein de la structure

Lors de mouvements de terrain, les déformations au sein de la structure peuvent être obtenues de deux manières différentes comme détaillé en introduction. L'approche fréquente consiste à découpler le problème et à appliquer les mouvements de sols déterminés en condition de terrain vierge directement aux fondations de la structure. Les effets de l'interaction sol-structure sont alors négligés.

Une approche plus complète consiste à considérer le problème comme un ensemble complexe, où l'on accorde la même importance aux phénomènes d'interaction sol-structure qu'au comportement du sol et à celui de la structure.

La figure 18 présente les déformations au niveau d'un certain nombre de points de la structure (figure 19) pour ces deux approches. Le choix des points géométriques de calcul des déformations a été guidé par la recherche des zones soumises aux déformations maximales lors de la conception de modèle réduit physique de structure (instrumentation par jauges de déformation). Par abus de langage, les valeurs expérimentales ainsi que les valeurs obtenues numériquement seront désignées par le même vocable "mesure de déformation" assignée à un lieu dénommé "jauge" (jauge A, jauge B, *etc.*).

Pour les simulations en condition de terrain vierge, ce sont des fuseaux de déformations qui sont tracés. En effet, les mouvements (deux translations et une rotation) de chaque semelle, sont affectés d'une certaine incertitude dépendant de la précision de la mesure et de la variabilité des résultats entre les différentes simulations. L'incertitude sur les translations est de l'ordre d'un millimètre, celle sur la rotation de 2°. Un grand nombre de simulations avec un modèle numérique de structure identique à celui utilisé dans la modélisation couplée sont réalisées en considérant un tirage aléatoire des incertitudes affectant les déplacements des semelles.

Chaque fuseau est alors défini par sa valeur moyenne, ses valeurs extrêmes (barres horizontales) et la plage centrale déterminée par la valeur moyenne plus ou moins l'écart-type. Les valeurs correspondant aux deux simulations en présence de la structure sont directement issues de l'état final de déformations au sein de la structure.

Les déformations obtenues en prenant en compte l'interaction sol-structure montrent certaines différences par rapport aux valeurs déterminées en traitant les mouvements de sol séparément du comportement de la structure. Certaines mesures de déformation présentent une augmentation de la valeur obtenue (jauge A, H et I), alors que pour les autres, la tendance est plutôt à la stagnation, ou à une très légère diminution (jauge F).

Ceci est différent des résultats des essais expérimentaux, où l'on obtenait de manière générale une réduction des déformations au sein de la structure, voire une inversion du signe des déformations pour les jauges H et I. La cohérence des résultats des simulations numériques avec ceux des essais expérimentaux antérieurs est donc assez moyenne, la raison étant les écarts de comportement de la partie FLAC$^{2D}$, ainsi que la solution technique retenue pour réaliser l'interfaçage de la structure avec les particules PFC$^{2D}$. Le faible nombre de simulations qui ont pu être réalisées en présence d'une structure en surface n'est pas à écarter et ne permet pas de se prononcer plus précisément.

## 5. Conclusion et perspectives

Un modèle numérique alliant deux approches complémentaires (Mécanique des Milieux Continus et Discontinus via FLAC$^{2D}$ et PFC$^{2D}$) a été développé afin de permettre l'étude des phénomènes d'interaction sol-structure lors de mouvements de terrain. Il a été utilisé dans le cadre particulier de l'effondrement d'une cavité donnant naissance à un fontis. Les résultats obtenus ont été comparés avec ceux antérieurs provenant d'essais expérimentaux sur modèle réduit. Bien que présentant certaines différences, les résultats des deux modèles convergent vers le fait qu'il n'est pas souhaitable de découpler la partie mouvement de terrains de celle concernant le comportement de la structure lors d'une telle étude. La prise en compte de l'interaction sol-structure permet en effet d'observer une redistribution des déformations dans la structure. Les déformations résultantes correspondent à une sous-évaluation des déformations subies en certains endroits, ce qui peut menacer la pérennité du bâtiment.

Le couplage FLAC2D-PFC2D s'est révélé pertinent. En effet, avec cet outil, il est maintenant possible de traiter des cas où l'on ne sait pas a priori ce qui se passe en surface. Cette approche améliore donc ce qui a pu être proposé par ailleurs, basé principalement sur une approche en milieu continu bien adaptée aux tassements induits par le creusement de tunnels. Même si les résultats et la conclusion a déjà été trouvé par d'autres (dans le domaine des tunnels principalement), l'apport de la modélisation par PFC2D est important dans la représentation du phénomène de rupture de la cavité (chute de blocs, foisonnement), la propagation d'un vide jusqu'en surface et la possible apparition de discontinuités importantes au sein des mouvements de terrain en étant capable de prendre en compte la présence d'une structure.

Les axes de développements futurs s'orientent à court terme vers la réalisation de nouvelles simulations en présence d'un bâtiment afin de confirmer ou d'infirmer les résultats obtenus. Cette étape requiert une amélioration préalable du modèle numérique couplé au niveau du comportement de la partie FLAC$^{2D}$.

A moyen terme, la réalisation d'études paramétriques permettant de caractériser l'influence de différents paramètres tels que la position de la structure, l'épaisseur et la constitution du recouvrement est envisagée.

## REMERCIEMENTS

## BIBLIOGRAPHIE

Abbass-Fayad, A. 2004. Modélisation numérique et analytique de la montée de cloche des carriers à faible profondeur. Etude de l'interaction sol-structure due aux mouvements de terrain induits par des fontis. Thèse de Doctorat INPL, 146p.

AFTES. (1995). "Recommandations relatives aux tassements liés au creusement des ouvrages en souterrain." Tunnels et Ouvrages Souterrains(132).

Augarde, C., Burd, H. & Houlsby, G. 1998. Some experiences of modelling tunnelling in soft gound using three-dimensional elements in A. Cividini (ed.), *Proceeding of the fourth European Conference on Numerical Methods in Geotechnical Engineering*, Udine. Wien:Springer-Verlag, 603-613.

Burland J. & Wroth C.. 1974. Settlement of buildings and associated damage. *State-of-the-Art Report for Session 5. Conf. on Settlement of Structures*, Cambridge, pp 611-654

Burland J. 1995. Invited Special Lecture: Assessment of risk of damage to buildings due to tunneling and excavation. 1st Int. Conf. on Earthquake Geotechnical Engineering, Tokyo, Vol 3; 1189-1201

Caudron M., Emeriault F., Kastner R. & Al Heib M. 2006a. Sinkhole and soil-structure interactions: Development of an experimental model. *Int. Conf. on Physical Modeling in Geotechnics, Hong-Kong*, 04-06 Aug. 2006, pp 1261-1267.

Caudron M., Emeriault F. & Al Heib M. 2006b. Numerical modeling of the soil-structure interaction during sinkholes. *Numerical Methods in Geotechnical Engineering, Graz*, 06-08 September 2006, pp 267-273.

Deck O., Al Heib M., and Homand F. 2003. Taking the soil-structure interaction into account in assessing the loading of a structure in a mining area. *Engineering Structure* 25:435-448.

Dolzhenko, N. & Mathieu P. 2001. Experimental modeling of urban underground works. Displacement measurement in a two-dimensional tunneling experiment. *Regional Conference on Geotechnical Aspects of Underground Construction in Soft Ground, Shangai*, 2001: 351-359.

Franzius J.N., Potts D.M., Addenbrooke T.I & Burland J.B 2004. The influence of building self-weight on tunnelling-induced ground and building deformation, *Soils and Foundations*, Vol. 44, No. 1, (2004), 25-38

ITASCA Consulting Group 2005. PFC2D 3.1 Manual provided with the PFC2D code.

Lake L.M., Rankin W.J. and Hawley J. 1992. Prediction and effects of ground movements caused by tunnelling in soft ground beneath urban areas. CIRIA funders report CP/5 129p.

Nakai, T.; Xu, L. & Yamazaki, H. 1997. 3D and 2D model tests and numerical analyses of settlements and earth pressures due to tunnel excavation, *Soils and Foundations*, 1997, 37, pp 31-42.

Pacyna, D.; Thimus, J. & Welter, P. 2005. Les carrières souterraines abandonnées en Belgique : impacts sur les infrastructures publiques et privées, *Actes des journées scientifiques du LCPC : Evaluation et gestion des risques liés aux carrières souterraines abandonnées*, mai 2005, pp 182-194

Peck, R. 1969. Deep excavations and tunnelling in soft ground, *Proceedings of the 7th International Conference on Soil Mechanics Foundation Engineering, Mexico*, 3:225-290.

Potyondy D.O., Cundall P.A. 2004. A bonded-particle model for rock. *International Journal of Rock Mechanics & Mining Sciences*, 41:1329-1364.

Vaillant J-M.; Mroueh H.; Shahrour I. 2005. Prise en compte de l`interaction sol-fondations-structures dans le calcul d`une structure flexible sur sol meuble. *Revue Européenne de Génie Civil (REGC), Vol.*9 No. 4, (2005). 497 – 508.

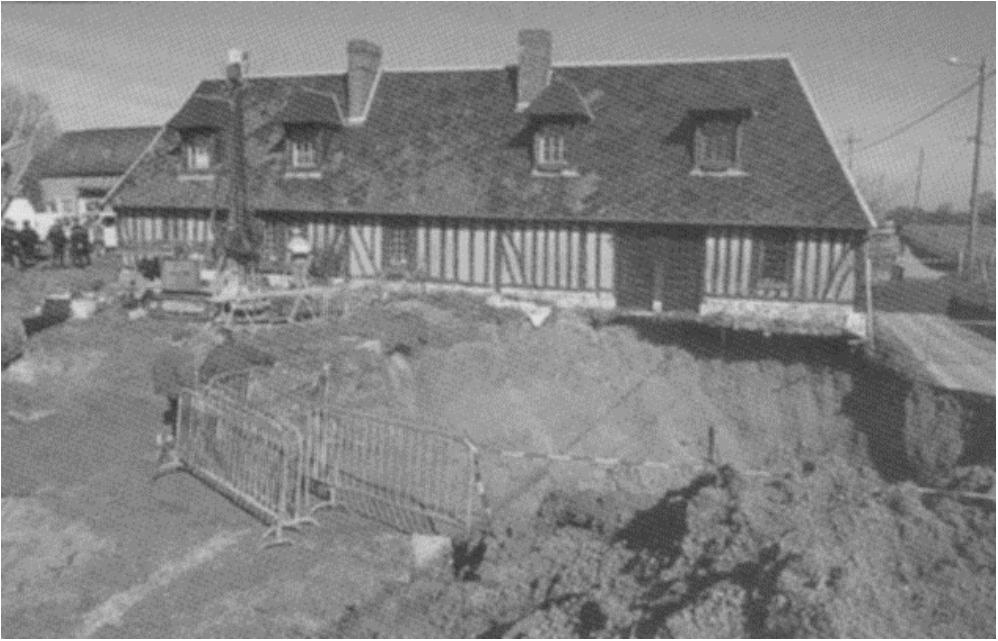

**Figure 1. Fontis de Neuville-sur-Authou (2001).**

Figure 1. Neuville-sur-Authou sinkhole (2001).

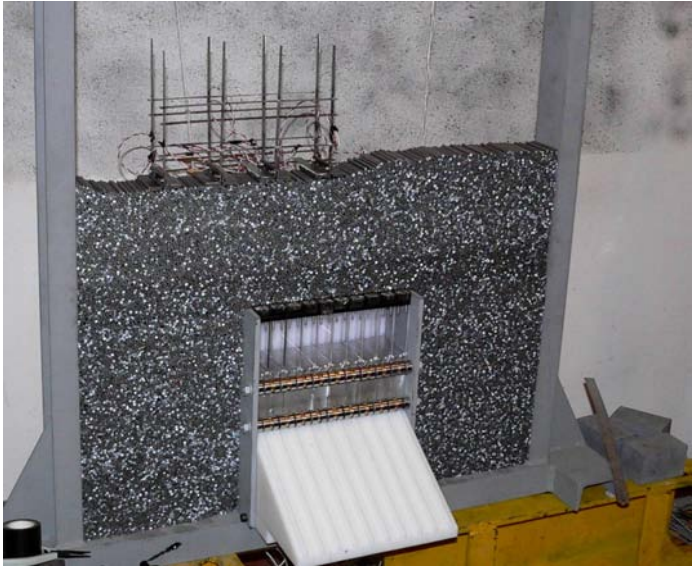

Figure 2. Photo du modèle réduit expérimental (dispositif de création de la cavité en bas au centre, massif de rouleaux de Schneebeli, maquette de bâtiment en surface).

Figure 2. View of the small-scale experimental model (apparatus for the creation of the cavity at the centred bottom, Schneebeli material, building model on the ground).

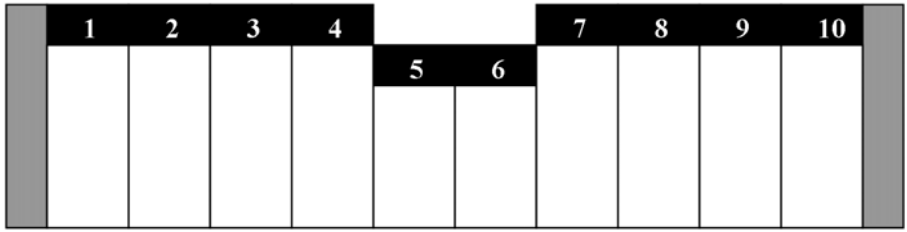

Figure 3. Dispositif expérimental de création de la cavité (les cales 5 et 6 sont abaissées, puis 4 et 7, etc.).

Figure 3. Experimental apparatus for the creation of the cavity (the moving parts 5 and 6 are lowered, and then number 4 and 7, and so on).

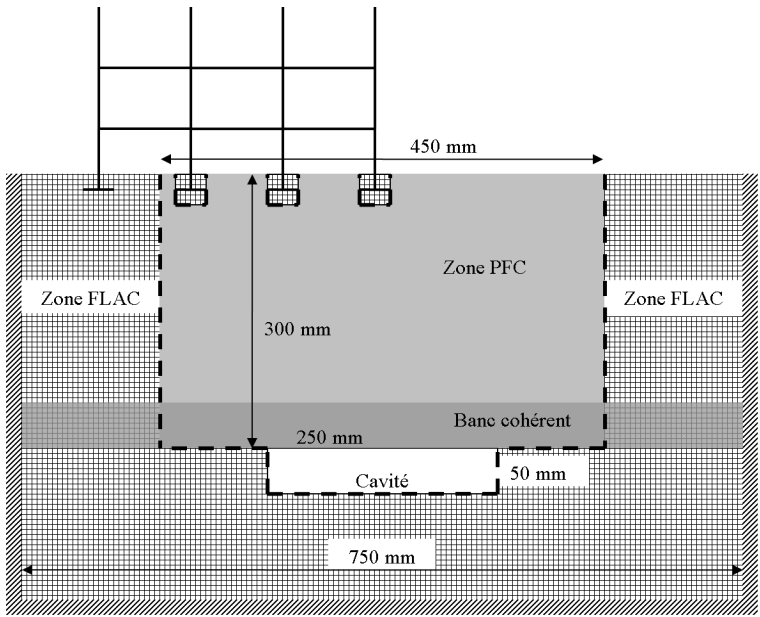

Figure 4. Coupe schématique du modèle numérique couplé FLAC$^{2D}$-PFC$^{2D}$.

Figure 4. Schematic cross section of the coupled model.

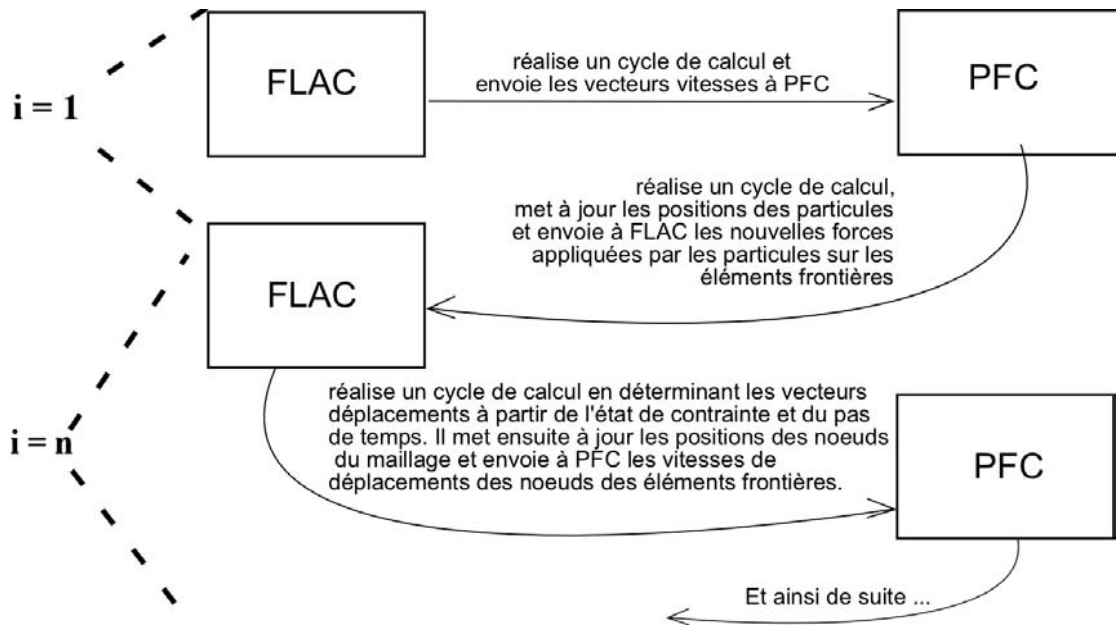

Figure 5. Illustration des principes du couplage entre FLAC$^{2D}$ et PFC$^{2D}$

Figure 5. Main principles for the coupling scheme between FLAC$^{2D}$ and PFC$^{2D}$.

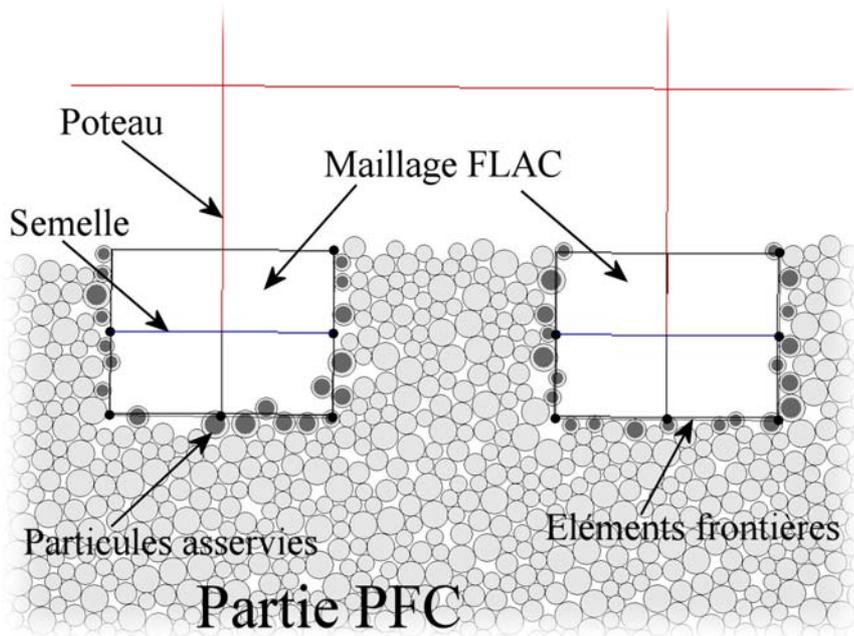

Figure 6. Solution retenue pour l'interfaçage des éléments de structure FLAC$^{2D}$ avec les particules PFC$^{2D}$.

Figure 6. Used solution for the coupling of FLAC$^{2D}$ structure elements and PFC$^{2D}$ particles.

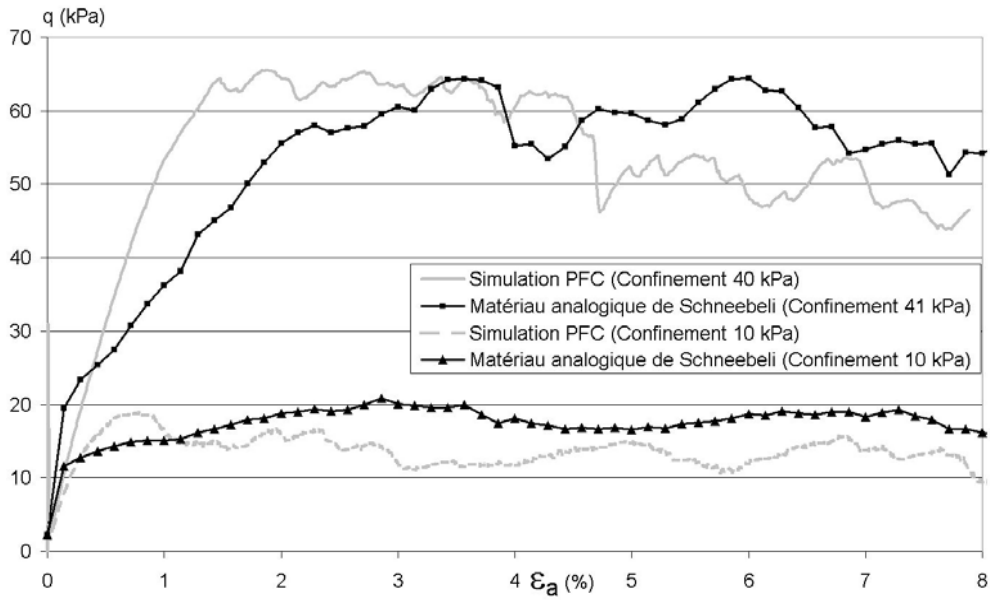

Figure 7. Essai biaxial - Courbe contrainte-déformation (q déviateur des contraintes – $\varepsilon_a$ déformation axiale).

Figure 7. Biaxial test – Deviatoric stress versus axial strain curve (q déviatoric stress – $\varepsilon_a$ axial strain)

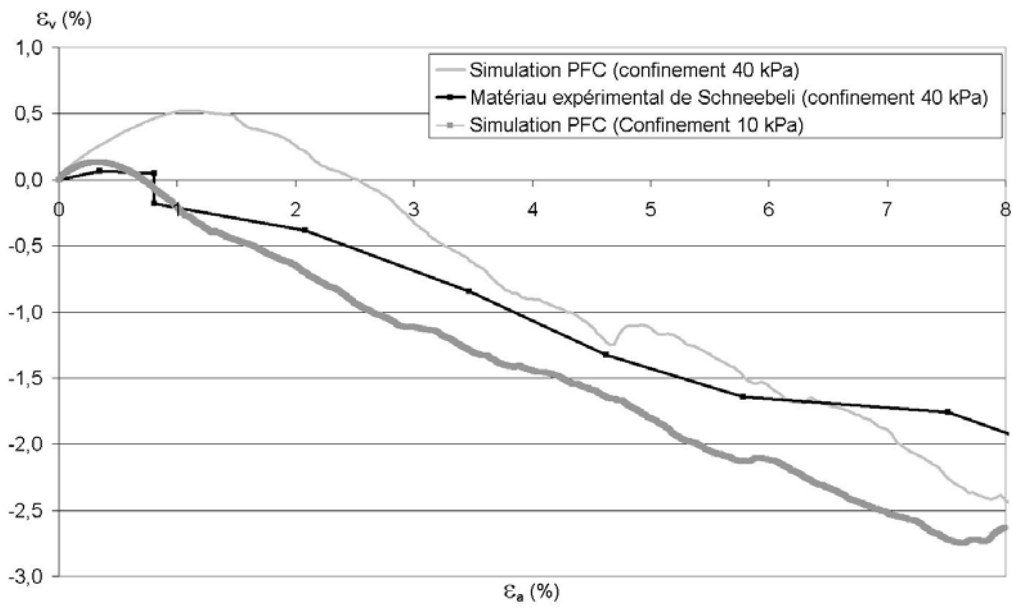

Figure 8. Essai biaxial – Courbes $\varepsilon_v$ déformation volumique- $\varepsilon_a$ déformation axiale.

Figure 8. Biaxial test – $\varepsilon_v$ Volumetric strain versus $\varepsilon_a$ axial strain curves.

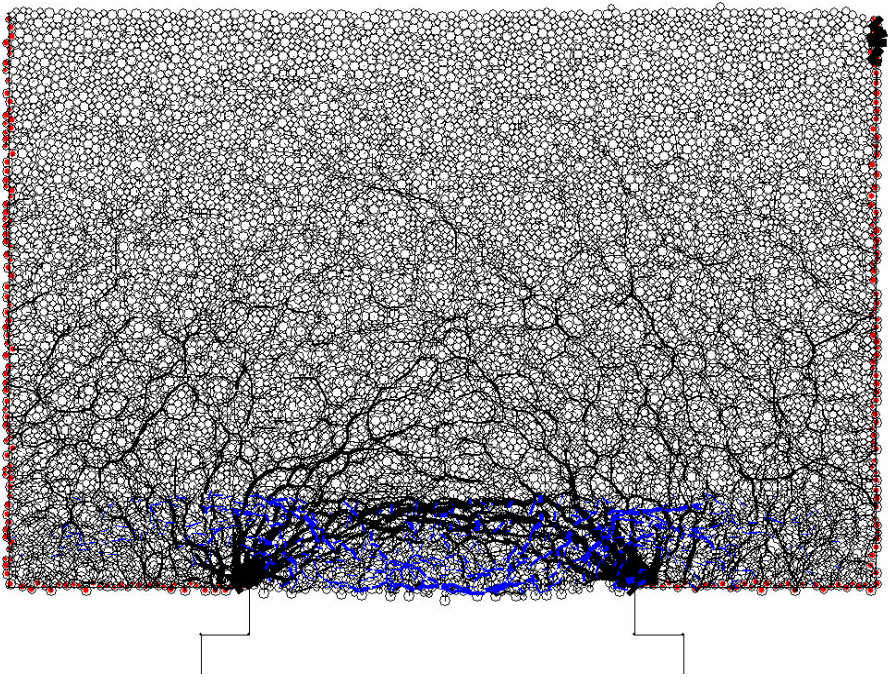

Figure 9. Quatrième et avant-dernière étape de création de la cavité. Les forces de contact et les efforts repris par les liaisons inter-particulaires sont tracés.

Figure 9. Fourth stage of the creation of the cavity. Contact forces and bonds forces acting on particles are plotted.

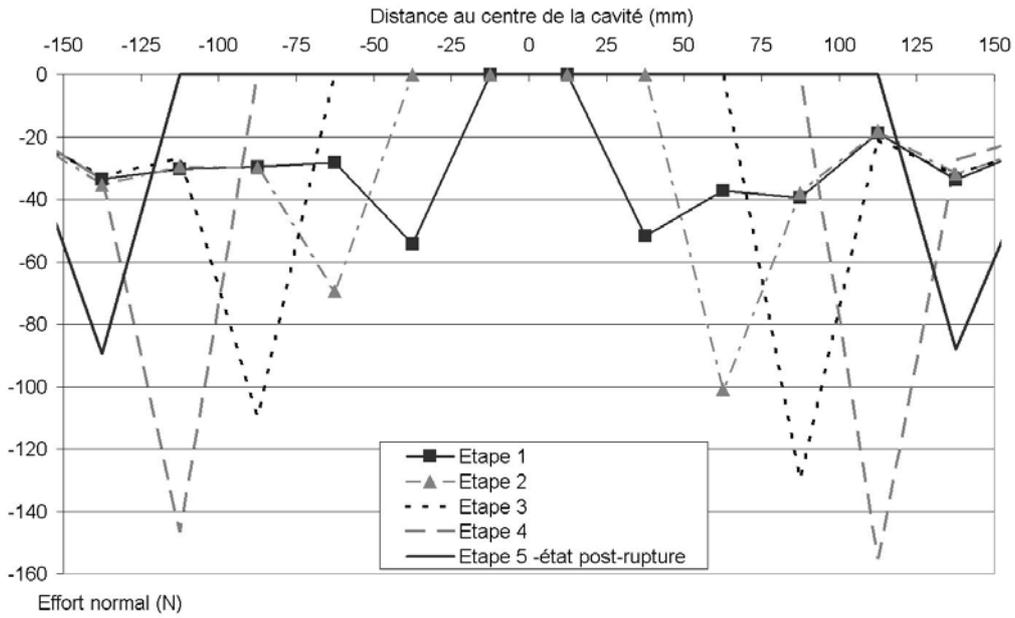

Figure 10a. Evolution des reports de charge sur les éléments frontières voisins lors de la création de la cavité pour la simulation 4.

Figure 10a. Evolution of the vertical forces acting at the level of the roof of the cavity during its creation for the fourth simulation.

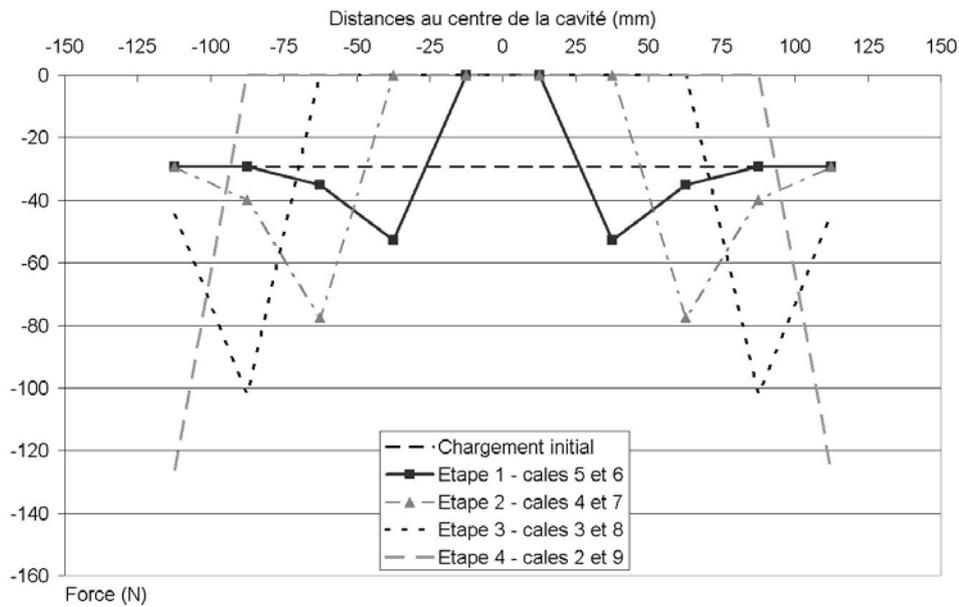

Figure 10b. Evolution des reports de charge sur les éléments frontières voisins lors de la création de la cavité (évolution théorique).

Figure 10b. Evolution of the vertical forces acting at the level of the roof of the cavity during its creation (theorical evolution).

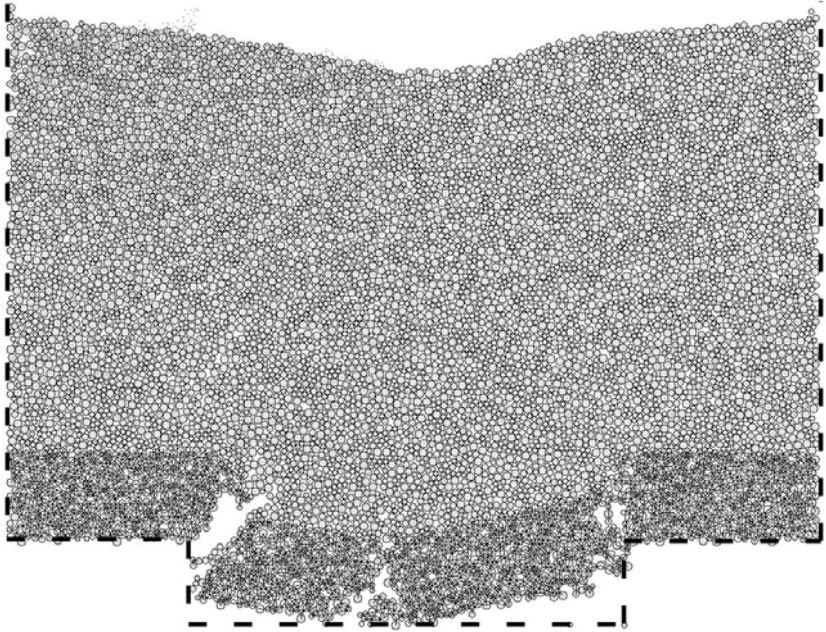

Figure 11. Etat final obtenu pour la quatrième simulation en condition de terrain vierge.

Figure 11. Final state corresponding to the fourth simulation in greenfield condition.

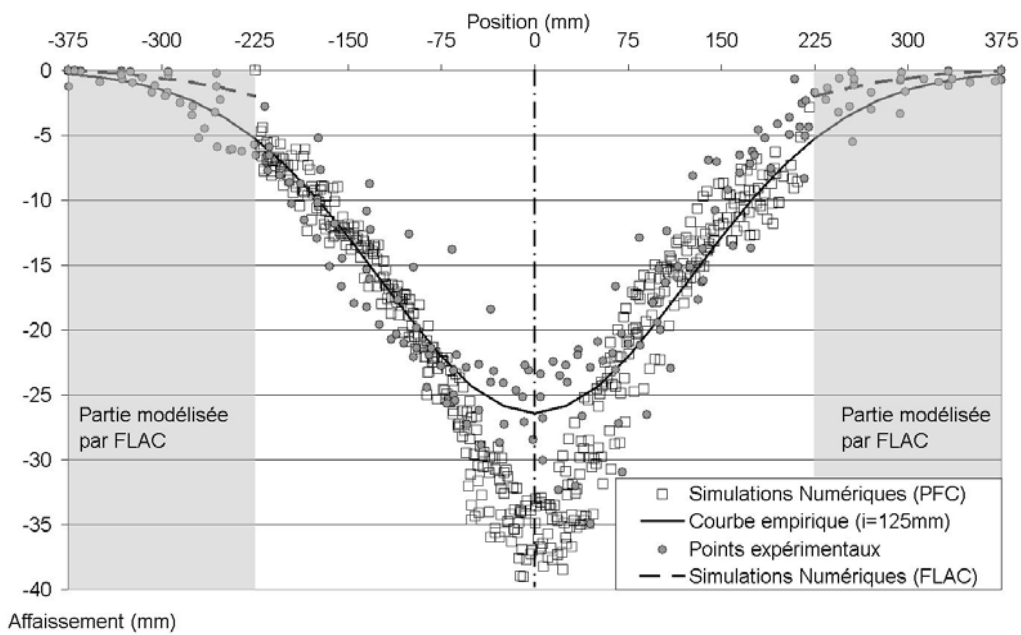

Figure 12. Cuvettes d'affaissement en surface en condition de terrain vierge (sont tracés les résultats expérimentaux, la courbe empirique calée sur ces résultats et les déplacements obtenus lors des cinq simulations).

Figure 12. Subsidence troughs in greenfield condition (experimental results, empirical approach and numerical results are plotted).

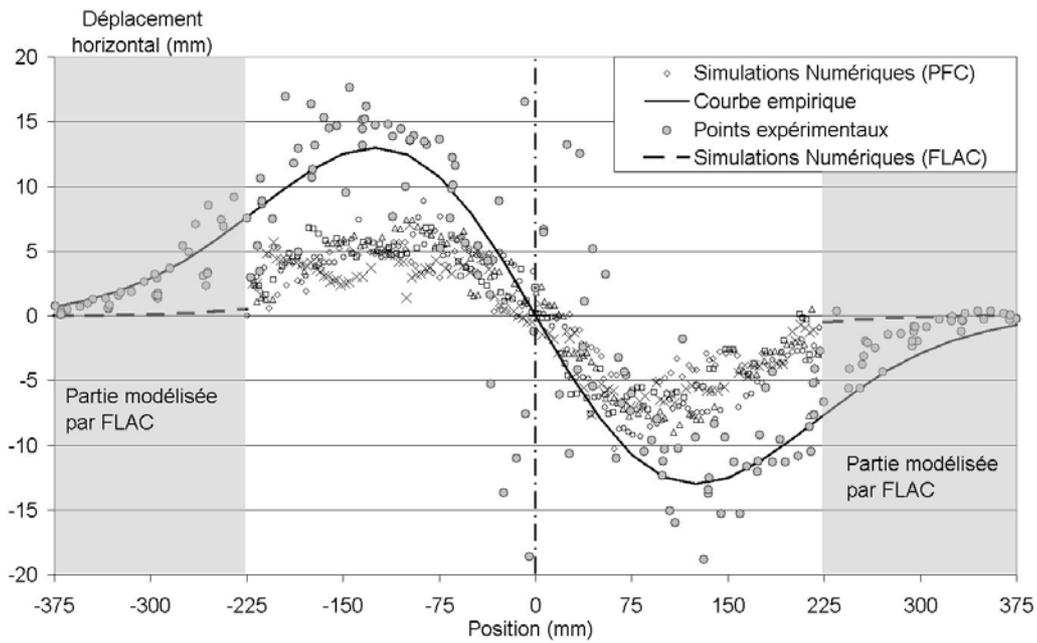

Figure 13. Déplacements horizontaux en surface en condition de terrain vierge (sont tracés les résultats expérimentaux, la courbe empirique calée sur ces résultats et les déplacements obtenus lors des cinq simulations).

Figure 13. Horizontal displacements in greenfield condition (experimental results, empirical curves and numerical results).

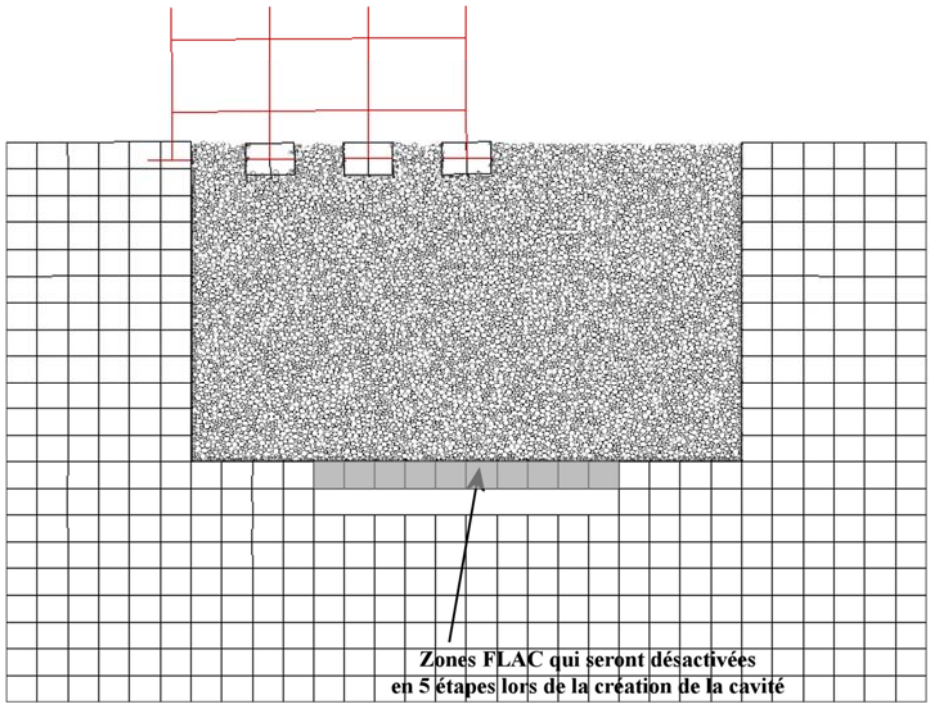

Figure 14. Capture de l'état initial d'une simulation réalisée avec le modèle de structure en surface.

Figure 14. Initial state corresponding to a computation with the building model on the ground.

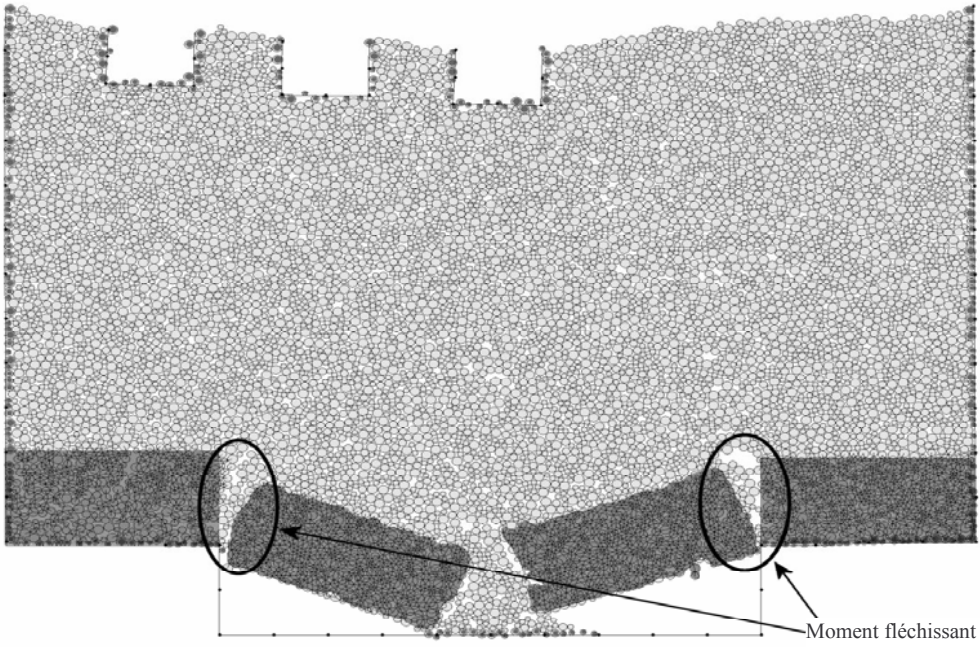

Figure 15. Détails de la partie PFC$^{2D}$ de la seconde simulation en présence de la structure après la rupture du banc résistant.

Figure 15. Details of the failure of the stiff bench corresponding to the second computation with the building model.

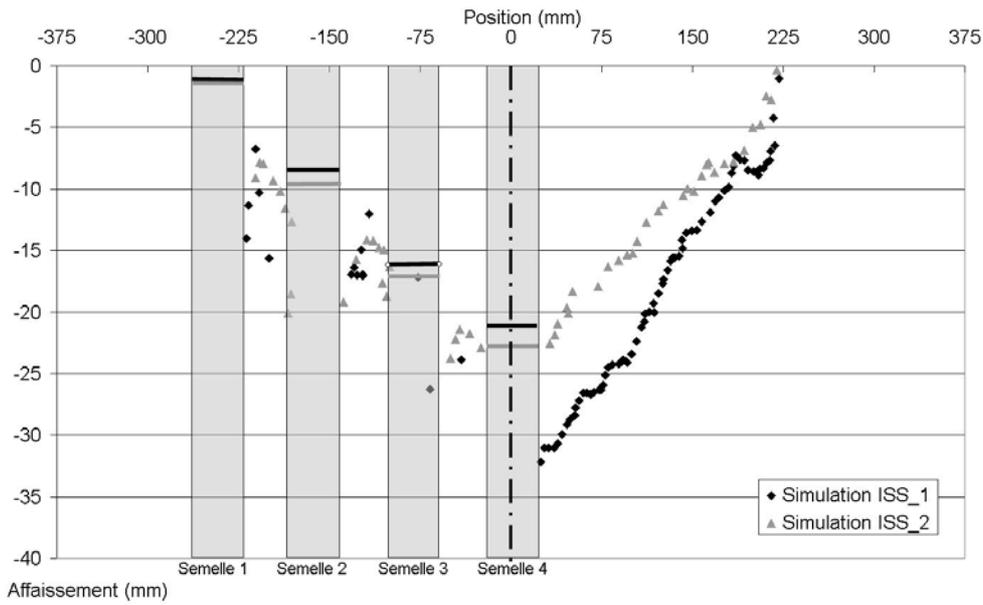

Figure 16. Cuvette d'affaissement en surface pour les deux simulations en présence de la structure en surface (les déplacements des semelles sont représentés par des segments).

Figure 16. Subsidence troughs for both computations with the soil-structure interaction (foundations displacements are plotted with a segment).

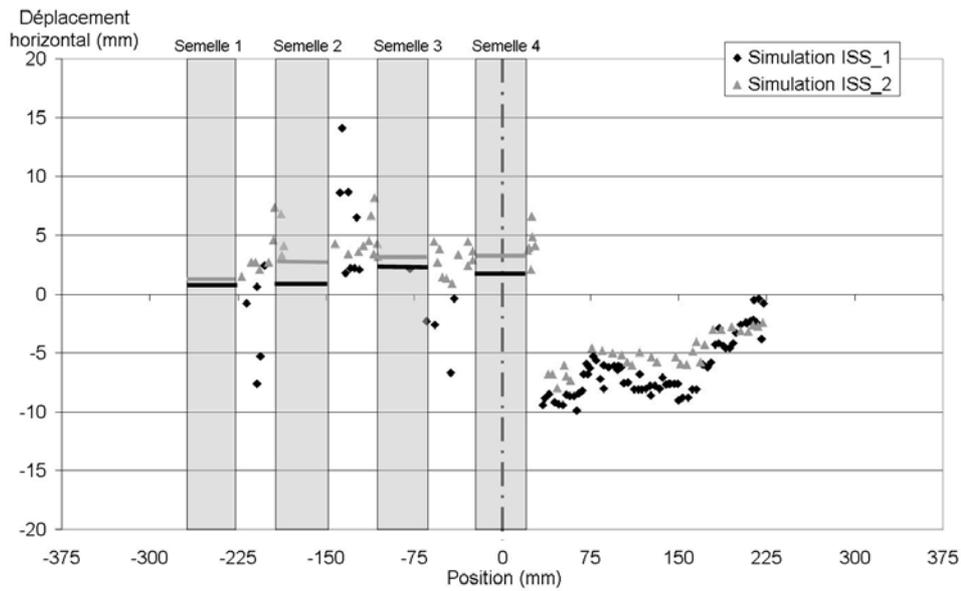

Figure 17. Déplacements horizontaux en surface en présence de la structure (les déplacements des semelles sont représentés par des segments).

Figure 17. Horizontal displacements for the computations with the soil-structure interaction (foundations displacements are plotted with a segment).

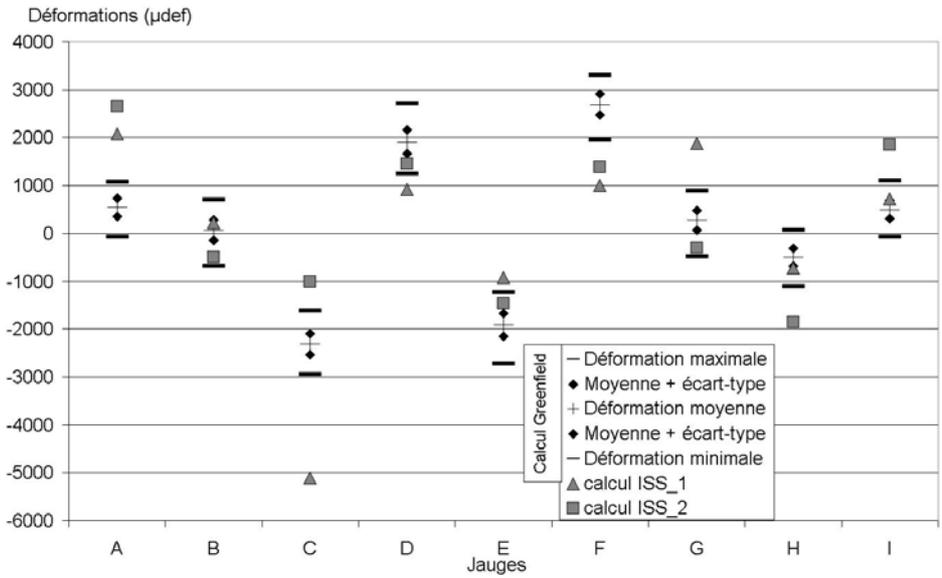

Figure 18. Déformations au sein de la structure lors du fontis (estimées pour l'approche en condition de terrain vierge).

Figure 18. Strains in the building caused by the sinkhole (estimated in case of the greenfield approach).

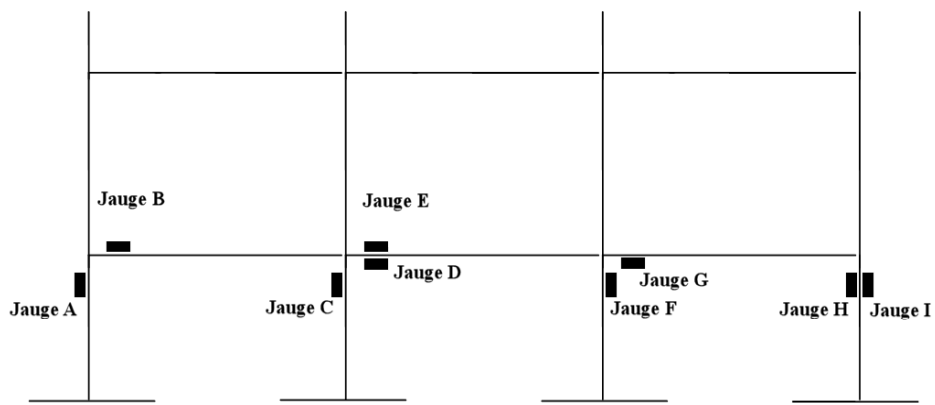

Figure 19. Localisation des jauges de déformation sur la structure.

Figure 19. Position of the strain gauges on the building model.